\documentstyle[epsf,prl,aps,twocolumn]{revtex}
\begin{document}
\draft
\title
{ Comments on tails in Schwarzschild spacetimes.}
\author
{Janusz Karkowski$^*$, Zdobys\l aw \'Swierczy\'nski$^{+}$ and
Edward Malec$^{*}$}
\address{ $^*$ Institute of Physics,  Jagiellonian University,
30-059  Cracow, Reymonta 4, Poland.}
\address{$^{+}$ Pedagogical University, Cracow, Podchor\c a\.zych 1, Poland.}
\maketitle
\begin{abstract}

We performed a careful numerical analysis of the late tail behaviour
of waves propagating in the Schwarzschild spacetime. Specifically the scalar
monopole, the electromagnetic dipole and the gravitational axial quadrupole
waves have been investigated. The obtained results agree with 
a falloff $1/t^{2l+3}$ for the
general initial data and $1/t^{2l+4}$ for the initially static  data.

\end{abstract}

\section{Introduction}

Waves propagating in a curved spacetime usually undergo backscatter;
that leads to the emergence of  two classes of effects, the so-called
quasinormal modes (discovered by Vishveshvara \cite{Vishveshvara})
 and the tails. Price, who  investigated the temporal
behaviour of tails in 1972, has shown some kind of
universality. The late tail behaviour of scalar, electromagnetic
and gravitational waves happened to depend   on the order of the multipole
expansion \cite{Price}. The exposition of \cite{Price} is somewhat
confusing and that forces us to say more than would normally be needed for the
purpose of a paper reporting only numerical results.

Price states    {\it If a static
l-pole field is present outside the star, prior to the onset of collapse,
the field will fall off as $t^{-(2l+2)}$. If there is no field
initially outside the star, but an l-pole perturbation develops
during the collapse process, it will fall off as $t^{-(2l+3)}$ at large
time \cite{Price}.}   That requires a comment.
  Since the evolution equation (\ref{2}) (see below) is linear and of
the form $-\partial_t^2\Psi +L\Psi =0$,
where $L$ is a linear time-independent operator, its solutions can be superposed.
That means that if there is a static solution, then it  cannot influence
any evolving perturbation. Irrespective of whether there are static solutions or
not, the perturbation will decay in the same way.
Therefore the quoted statement  cannot be taken at the face value. The
natural understanding  would be the following:
{\it A static field }   means that
there is initially a field ($\Psi (t=0)\ne 0$), but $\partial_t\Psi =0$ (hence data are
momentarily  static).   {\it The initial absence of field} in turn  means
that  $\Psi =0$ but $\partial_t\Psi \ne 0$.
This agrees with the standard way of posing the initial
value problem in mathematical literature, and with this in mind we
inspect  the later papers on this subject.
The notion of {\it initially static l-pole} has been clearly specified
in two publications that appeared in late seventies; it becomes
 an {\it initially stationary multipole of order l} (that falls off
like $1/t^{2l+2}$) \cite{Moncrief}.  Therein
the term {\it initially stationary} means {momentarily static} as defined above,
 but with $\Psi $ satisfying special asymptotic condition.
 On the other hand a detailed discussion   in \cite{Leaver}  shows that for
{\it initially nonstatic, $\partial_t\Psi \ne 0$ } analytic initial data  satisfying
 certain boundary conditions (at the event  horizon of a black hole and
 at spatial infinity) the decay  of a solution agrees with that of the Price's
 {\it initially static l-pole field}. This might suggest in turn that
  one can   understand the "static l-pole field" as a specially
prepared {\it general} initial wave profile.  Corresponding initial data
(both $\Psi $ and $\partial_t\Psi $ are nonzero)
   have a noncompact support (the spatial support has infinite extension, as measured in
 terms of the tortoise variable $r^*$ -- see below).
The work of Gundlach et al. \cite{Gundlach}  numerically confirmed the 
analytically derived $1/t^{2l+2}$ decay of late tails.
Let us remark an internal inconsistency -- the contention must be now that 
the fall off  actually   depends on the profile of initial data,
in contrast to the  primary expectation that
the   "tail" is a universal phenomenon due to the backscatter of the 
wave signal off the curvature of the geometry.
 
Leaving aside the     area of {\it initially static fields},
let us  notice   that in the case of   data of compact support the
predictions of \cite{Price} are clear -- solutions should have a   falloff    $t^{-(2l+3)}$.
That estimate applies to solutions generated by
{\it generic initial data} ($\Psi \ne 0$, $\partial_t\Psi \ne 0$) \cite{Leaver}.
From a naive inspection of the formal solution $\Psi =\partial_tG*\Psi (t=0)+
G*\partial_t\Psi (t=0)$ (where $G$  is the Green function), one expects that
in the case of momentarily static initial data the fall off is
faster (by $1/t$) than that corresponding to general initial data.
It is interesting  that  no numerical investigation has been performed
in the case of moment of time symmetry initial data of compact support.
It should be noticed also that
the status of the analytic investigation is still not quite satisfactory,
since it   consists either of heuristic
analysis \cite{Price} or rather formal investigation of the Green function
formulae (\cite{Leaver}, \cite{Ching})  \cite{all}. It is only
recently that the (generic case) falloff $1/t^{ 3}$
has been  rigorously proven by Dafermos and Rodnianski for the monopole
mode of  the real  scalar field \cite{Dafermos}.
The same    conclusion
is being derived, in a different (spectral) method, by Machedon and Stalker
\cite{Stalker}.

The main focus of our numerical work will be on
finding the falloff of tails generated by   initial data of compact support in either
of the aforementioned two fundamental cases.
  We study the  evolution of the scalar monopole ($l=0$),
the electromagnetic  dipole ($l=1$) and the gravitational axial quadrupole
($l=2$).  The obtained results reveal  a falloff $1/t^{2l+3}$ for the data
$\partial_t\Psi_l\ne 0, \Psi_l=0$, in agreement with \cite{Price},
and $1/t^{2l+4}$ for the initially static  initial data \cite{remark}.  This latter
result on the behaviour of late  tails generated by initial data
$\partial_t\Psi_l=0, \Psi_l\ne 0$   is new in the numerical
literature.
Thus the generic initial data have the falloff $1/t^{2l+3}$, which
agrees with Leaver \cite{Leaver}, Ching et al. \cite{Ching} and
 Poisson \cite{Poisson}.

Our results support the view    that   the l-th moment of
any of the waves (scalar, electromagnetic or gravitational) will decay like
$1/t^{2l+3}$  for general initial data (in accordance with \cite{Ching}, \cite{Leaver}
 and \cite{Poisson}) and $1/t^{2l+4}$ for the l-th moments evolving from
initially static    data   (in agreement with the formal analysis
of  \cite{Leaver} and \cite{Poisson}).
These results should be of practical significance for numerical relativity. The determination of
the late tail behaviour is a   nontrivial but at the same time feasible numerical task
that can serve as a useful test for the  accuracy and stability of numerical codes.

\section{Definitions and equations}

  The spacetime geometry is defined by the line element
\begin{equation}
ds^2=-\eta_Rdt^2+{dR^2\over \eta_R}+R^2d\Omega^2,
\label{1}
\end{equation}
where $t$ is a time coordinate, $R$ is the radial areal coordinate,
$\eta_R=1-{2m\over R}$ and $d\Omega^2=d\theta^2+\sin^2\theta d\phi^2$
is the line element on the unit sphere, $0\le \phi < 2\pi $ and
$0\le \theta \le \pi $. Throughout this paper the Newtonian constant $G$ and
the velocity of light $c$ are put equal to 1.

The propagation of the scalar, the dipole electromagnetic and
the axial (quadrupole)  gravitational waves is given by
\begin{equation}
(-\partial_t^2+\partial_{r^*}^2)\Psi = V\Psi .
\label{2}
\end{equation}
Here $r^*(R)=R+2m\ln \Bigl( {R\over 2m}-1\Bigr) $ is the tortoise coordinate
while the potential term reads:
for the $l=0$ mode of the scalar field
\begin{equation}
V(R)=2m{\eta_R\over R^3},
\label{3}
\end{equation}
for the $l=1$  (dipole) electromagnetic mode
\begin{equation}
V(R)=2{\eta_R\over R^2}
\label{4}
\end{equation}
and for the quadrupole axial   mode of  GW
\begin{equation}
V(R)=6{\eta^2_R\over R^2} (1-{m\over R}).
\label{5}
\end{equation}

\section{Numerical schemes}

We will choose following classes of  initial data:

i) $\Psi (R,t=0)= \sin^4\Bigl( \pi {r^*-r^*(a)\over 4m}\Bigr) $ and
$\partial_t\Psi |_{t=0}=
-\partial_{r^*} \Psi_{t=0}$ for $r^*\in (r^*(a)=r^*(3m), r^*(a)+4m)$,
 and $\Psi (r,0)=\partial_t\Psi =0$
for $r^*<r^*(a) $ or $r^*>r^*(a)+4m$.
 The expected asymptotic tail falloff  should be  the same as
for generic initial data.

ii) Symmetric initial data with $\Psi (R,t=0)= \sin^4\Bigl( \pi {r^*-r^*(a)\over 4m}\Bigr) $
for $r^*\in (r^*(a)=r^*(3m), r^*(a)+4m)$  and   $\Psi (r,0)=0$
for $r^*<r^*(a) $ or $r^*>r^*(a)+4m$, and $\partial_t\Psi_{t=0}=0$ everywhere.
It is shown below that the corresponding tail  decays faster than that of i).

iii) $\partial_t\Psi =\sin^4\Bigl( \pi {r^*-r^*(a)\over 4m }\Bigr) $
for $r^*<r^*(a) =r^*(3m)$ or $r^*>r^*(a)+4m$ and $\partial_t\Psi (r,0)= 0$
for $r^*<r^*(a) $ or $r^*>r^*(a)+4m$, while $\Psi |_{t=0}=0$ everywhere;
 the tail behaviour happens to be  like in i).

One   finds convenient to  split the wave equation  (\ref{2}),
into the pair of first order differential equations
\begin{equation}
(\partial_t + \partial_{r^*}) \Psi = \Phi
\end{equation}
\begin{equation}
(\partial_t - \partial_{r^*}) \Phi = - V({r^*}) \Psi .
\label{8}
\end{equation}
We use the following mesh scheme    (see Fig. 1)
%
%
\begin{figure}[1]
\epsfxsize=6cm
\centerline{\epsffile{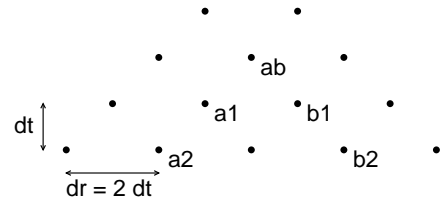}}
\caption{This figure shows the mesh used in the numerical computations. In the  
one-step scheme the functions $\Phi_{ab}, \Psi_{ab}$ are computed from
Eqs (\ref{9} -- \ref{10}) using values at points $a1$ and $b1$. The second
algorithm needs values at four points $a1,a2,b1$ and $b2$ in order to obtain
$\Phi_{ab}, \Psi_{ab}$   from Eqs (\ref{13}).}
\end{figure}
%
%
and two differencing schemes: the one-step implicit alghoritm
and the multistep Adams-like algorithm. In the first approach we approximate
the equations (\ref{8}) by
\begin{equation}
\Psi_{ab} - \Psi_{a1} = ( \Phi_{ab} + \Phi_{a1} ) \frac{dt}{2}
\end{equation}
\begin{equation}
\Phi_{ab} - \Phi_{b1} = - ( V_{ab} \Psi_{ab} + V_{b1} \Psi_{b1}) \frac{dt}{2}.
\label{9}
\end{equation}
In the next step these  equations are solved  for $ \Psi_{ab}, \Phi_{ab} $,
\begin{equation}
\Psi_{ab} = \Psi_{ab}( \Psi_{a1}, \Psi_{b1}, \Phi_{a1}, \Phi_{b1} )
\end{equation}
\begin{equation}
\Phi_{ab} = \Phi_{ab}( \Psi_{a1}, \Psi_{b1}, \Phi_{a1}, \Phi_{b1} ).
\label{10}
\end{equation}
The second alghoritm is based on the Adams-Moulton corrector
formula   for ordinary differential equations  \cite{Press}
\begin{equation}
y_{n+1} = y_{n} + \frac{dt}{12}
	(5y^{'}_{n+1} + 8y^{'}_{n} -y^{'}_{n-1}) + O(dt^{4})
        \label{11}
\end{equation}
Applying this formula to each of the equations (\ref{8}) we get approximatelly
\begin{equation}
\Psi_{ab} = \Psi_{a1} + \frac{dt}{12}
(5 \Phi_{ab} + 8 \Phi_{a1} - \Phi_{a2})
\label{12}
\end{equation}
and
\begin{equation}
\Phi_{ab} = \Phi_{b1} - \frac{dt}{12}
	    (5 V_{ab} \Psi_{ab} + 8 V_{b1} \Psi_{b1} - V_{b2} \Psi_{b2})
\label{13}
\end{equation}
Again this linear set of equations can be solved for $ \Psi_{ab}, \Phi_{ab} $
\begin{eqnarray}
\Psi_{ab} &=& \Psi_{ab}( \Psi_{a1}, \Psi_{b1}, \Psi_{a2}, \Psi_{b2},
			\Phi_{a1}, \Phi_{b1}, \Phi_{a2}, \Phi_{b2} )  \nonumber\\
&&\Phi_{ab} = \Phi_{ab}( \Psi_{a1}, \Psi_{b1}, \Psi_{a2}, \Psi_{b2},
			\Phi_{a1}, \Phi_{b1}, \Phi_{a2}, \Phi_{b2} ).\nonumber\\
\end{eqnarray}
Let us stress that these two alghoritms  give almost the same results,
hinting at  the numerical stability of our methods.

\section{Numerical results}

We expect the field $ \Psi_l(r^*,t) $   to behave (for a fixed $r^*$)
like
\begin{equation}
\lim_{t \rightarrow \infty } \Psi(r^*,t) = C t^{- \alpha}
\end{equation}
Our goal is to get a   value of $ \alpha $ with  a reasonable accuracy.
 Therefore we calculate the function
\begin{equation}
f(t) = - \frac {d \log(\Psi_l(r^*,t))}{d \log(t)} =
	- t \frac {d \log(\Psi_l(r^*,t))}{dt}
\end{equation}
which asymptotically should be equal to $ \alpha $.

In the first instance we analysed the  case with
  waves, which belongs to the class i) of Sec. 3, and
for the initial data of class iii) of the preceding section.
The results are presented in Figs. 2 -- 4 which show the temporal behaviour
 of the funtion
$f(t)$ as seen at $r^*=r^*(3m)+4m $.  In all figures (2 -- 7) below
the time is put on the x-axis, which  is scaled in units of $m$.
\begin{figure}[2]
\epsfxsize=6cm
\centerline{\epsffile{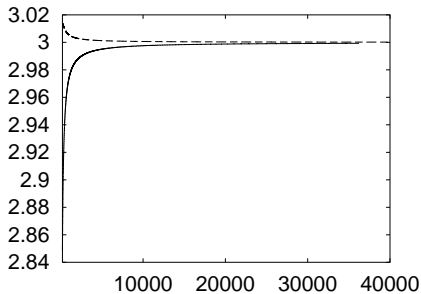 }}
\caption{
The solid (general initial data) and  broken line  (data with $\phi =0$),
 the behaviour of $f(t)$ for the scalar waves. }
\end{figure}
%
The numerical time ranged from   5000 m
(for the gravitational and electromagnetic waves) up to   40000 m (for the scalar waves).
It turned out that the precise shape of this function, without numerical noise,
can be calculated using numbers with 64 significant digits (gravitational  
and electromagnetic cases) and the ordinary double precision numbers
(scalar case). In order to achieve this aim we have used the freely distributable "qd"
(quad precision) and "arprec" (arbitrary precision) numerical libraries \cite{miller}.
%
\begin{figure}[3]
\epsfxsize=6cm
\centerline{\epsffile{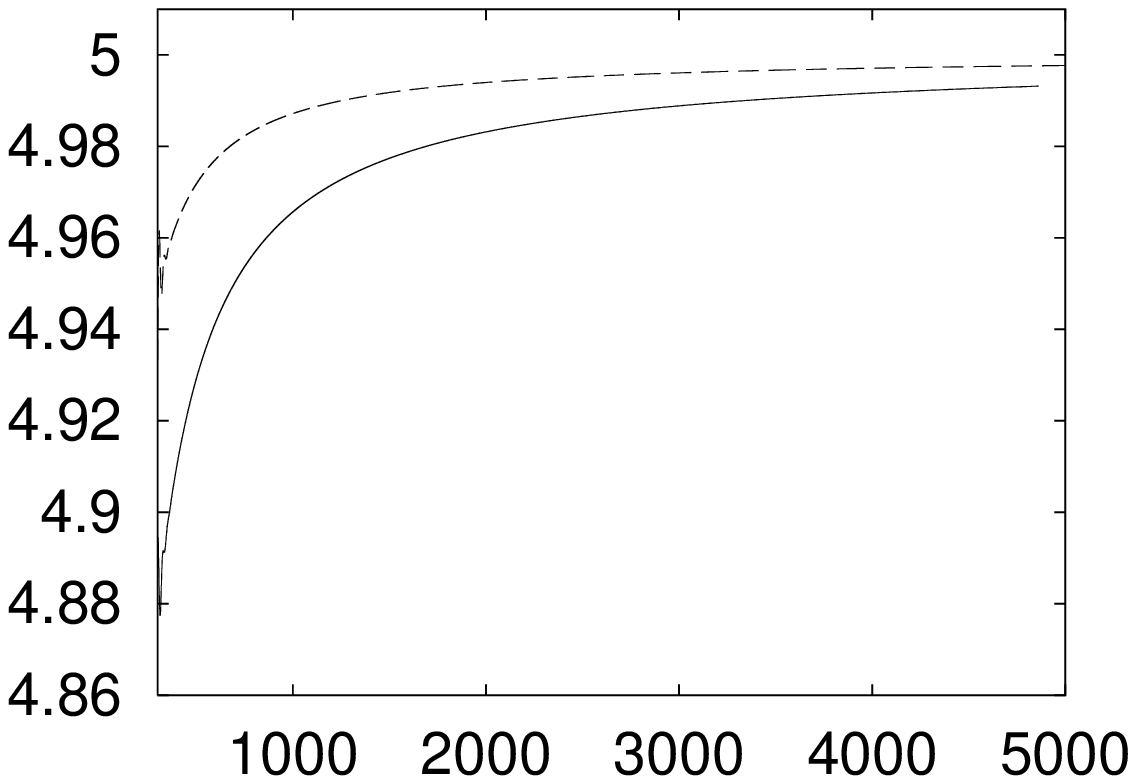 }}
\caption{
The solid (general initial data) and   broken line  (data with $\phi =0$),
 the behaviour of $f(t)$ for the selectromagnetic waves. }
\end{figure}
%
\begin{figure}[4]
\epsfxsize=6cm
\centerline{\epsffile{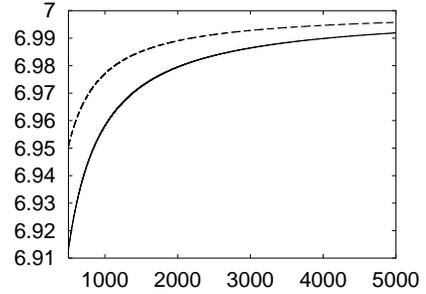 }}
\caption{
The solid (general initial data) and   broken line  (data with $\phi =0$),
 the behaviour of $f(t)$ for the gravitational waves. }
\end{figure}
 %
The high accuracy calculations are  time consuming and
therefore the achieved calculational  time   for the scalar case  exceeds by
 a factor  of 8 the evolution time for the remaining  cases.
The numerically achieved values of the  exponent , for the general initial data i),
range from  2.999  for the scalar field,
through 4.99  for the electromagnetic field up to 6.99  for the
gravitational waves (figures 2 -- 4, solid line).
  Both sets of exponents is in a good agreement with
 3, 5 and 7, respectively, obtained by Ching et al. \cite{Ching}.
   Similar results (broken line Fig. 3)
have been obtained for the initial data of class iii) of the preceding section,
in agreement with \cite{Price}.

The momentarily static initial data (class ii) from Sec. 3)   produced
profiles $f(t)$ shown in Figs. 5 -- 7.  In this case the numerical exponents have
been determined at $r^*(b)=r^*(3m)+4m$ and also at another
obervation point, $r^*=100m +r^*(b)  $. In the first case
we obtained 4, 6 and 8 (up to the fourth digit number)
while in the latter case we arrived at $3.99$, $5.88$
and $7.84$, for the scalar, electromagnetic and gravitational waves, respectively.
 These  exponents seem to be stabilize at  4, 6 and 8
for the scalar, electromagnetic and gravitational fields, respectively; notice that
the results detected   at $r^*=r^*(3m)+4m$ converge quicker to the prospective limiting value
than those taken at the  point located farther. The asymptotic tail regime is evidently
achieved quicker at a region closer to the horizon, which is a new feature, unknown
in the existing literature. But in both detection points, the asymptotic
values  of $\alpha $ are similar and   very close in the case of the scalar waves,
see Fig. 5.  The calculation of these exponents  constitutes the main
 result of this paper.

\begin{figure}[5]
\epsfxsize=6cm
\centerline{\epsffile{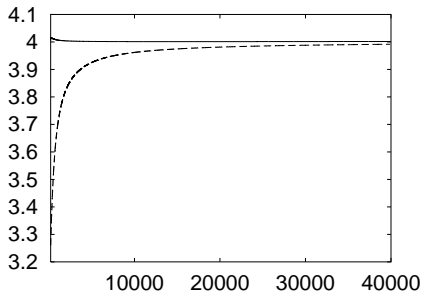 }}
\caption{ Time-symmetric initial data.
The solid and   broken lines  show the behaviour of $f(t)$ for
scalar  waves, as seen at the observation points
$r^*(b)= r^*(3m)+4m $ and $r^*=r^*(b)+100m$, respectively.   }
\end{figure}

\begin{figure}[6]
\epsfxsize=6cm
\centerline{\epsffile{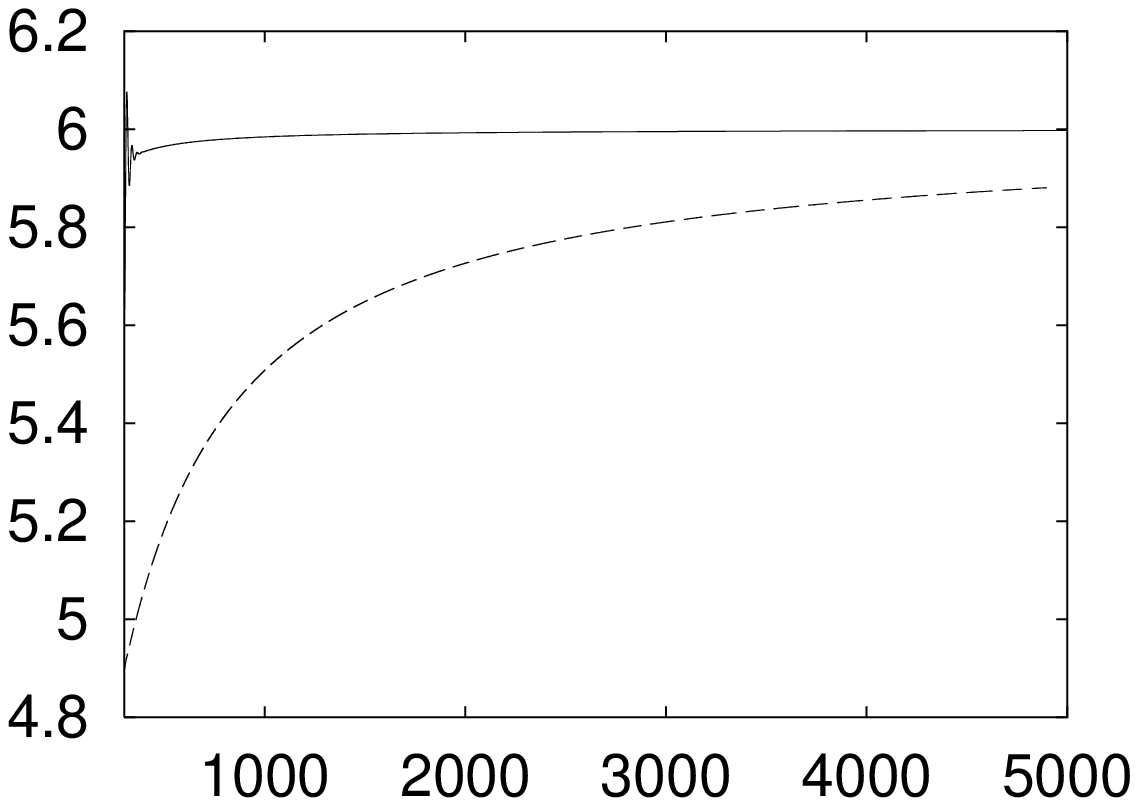 }}
\caption{ Time-symmetric initial data.
The solid and   broken lines  show the behaviour of $f(t)$ for
electromagnetic  waves, as seen at the observation points
$r^*(b)= r^*(3m)+4m $ and $r^*=r^*(b)+100m$, respectively.   }
\end{figure}

\begin{figure}[7]
\epsfxsize=6cm
\centerline{\epsffile{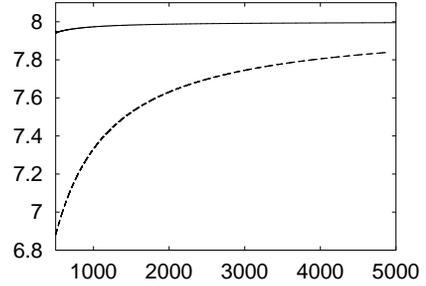 }}
\caption{ Time-symmetric initial data.
The solid and   broken lines  show the behaviour of $f(t)$ for
gravitational  waves, as seen at the observation points
$r^*(b)= r^*(3m)+4m $ and $r^*=r^*(b)+100m$, respectively.   }
\end{figure}
Acknowledgments.    This work has been supported
in part  by the KBN grant 2 PO3B  006 23. One of us (EM) acknowledges 
a discussion with Eric Poisson.

\end{document}